\RequirePackage{snapshot}
\documentclass[letterpaper,twocolumn,10ptl]{IEEEtran}

\usepackage{hyperref}
\usepackage{lineno}
\modulolinenumbers[5]

\usepackage[american]{babel}
\usepackage{inputenc}
\usepackage{graphicx}
\usepackage{verbatim}
\usepackage{bbm}
\usepackage[cmex10]{amsmath}
\usepackage{array}
\usepackage{paralist}
\usepackage{amsthm}
\usepackage{amssymb}
\usepackage{pgfplots}
\usepackage{tikz}
\usetikzlibrary{shapes}
\usetikzlibrary{plotmarks}
\usepackage[ruled]{algorithm2e}
\usepackage[caption=false]{subfig}
\usepackage{epstopdf}
\usepackage{hhline}
\usepackage{nicefrac}
\usepackage{dblfloatfix}

\newcommand{\cell}{\mathcal{C}}
\renewcommand{\Pr}{\mathbf{P}}
\newcommand{\Ex}{\mathbf{E}}
\newcommand{\ind}{\mathbf{1}}
\newcommand{\coverageD}{\mathcal{P}}
\newcommand{\content}{I}
\newcommand{\coverageN}{N}
\newcommand{\coverageTDF}{{\overline P}}
\newcommand{\popularity}{A}

\newcommand{\cache}{C}
\newcommand{\C}{C}
\newcommand{\hit}{\delta}
\newcommand{\hitP}{\mathbb{P}_{\mathrm{hit}}}
\newcommand{\contentset}{\mathbb{J}}
\newcommand{\coverageset}{\mathbb{N}_0}
\DeclareMathOperator*{\argmax}{arg\,max}
\newcommand{\calI}{\mathcal{I}}
\newcommand{\calJ}{\mathcal{J}}

\newtheorem{theorem}{Theorem}
\newtheorem{lemma}[theorem]{Lemma}
\newtheorem{proposition}[theorem]{Proposition}

\newtheorem{rem}[theorem]{Remark}
\newenvironment{remark}{\begin{rem}\rm}{\end{rem}}

\begin{document}

    \title{Optimal Geographic Caching in Cellular Networks with Linear
      Content Coding}

\author{
\IEEEauthorblockN{
Jocelyne Elias\IEEEauthorrefmark{1}
and
Bart{\l}omiej B{\l}aszczyszyn\IEEEauthorrefmark{2}
\thanks{
\IEEEauthorrefmark{1}Jocelyne Elias is with Inria/ENS, Paris and LIPADE Laboratory, Paris Descartes University, France. E-mail: jocelyne.elias@parisdescartes.fr.
Bart{\l}omiej B{\l}aszczyszyn is with Inria/ENS, Paris, France. E-mail: Bartek.Blaszczyszyn@ens.fr.
}
}
}

\maketitle

\begin{abstract}
We state and solve a problem of the optimal geographic caching
of content in cellular networks, where {\em linear combinations of contents} are stored in the caches of base
stations. 
We consider a general content popularity distribution and a general distribution of the number of stations covering the typical location in the network. We are looking for a policy of content caching maximizing the probability of serving the typical content request from the caches of covering stations.
The problem has a special form of {\em monotone sub-modular}
set function maximization.
Using {\em dynamic programming}, we find a deterministic policy solving the problem.
We also consider two natural {\em greedy caching policies}.
We evaluate our policies considering two popular stochastic geometric 
coverage models: the Boolean one and the Signal-to-Interference-and-Noise-Ratio one, assuming Zipf popularity distribution.
Our numerical results show that the proposed deterministic policies
are in general not worse than some randomized policy considered in the literature and can further improve the total hit probability in the moderately high coverage regime.
\end{abstract}

\begin{keywords}
Cellular caching, Network coding, Hit probability,  Coverage model,
Optimization, Stochastic Geometry.
\end{keywords}

\section{Introduction}
\label{introduction}

The rapid proliferation of smartphones, tablets and other smart mobile devices over the last few years has come hand in hand with content-oriented services, which are actually dominating the Internet traffic. According to~\cite{cisco2016mobiledata}, video streaming for example accounts for 54\% of the total Internet traffic, and the ratio is expected to grow to 71\% by the end of 2019.
This phenomenon has posed new challenges for mobile network operators and has pushed them to implement novel schemes to efficiently operate their cellular infrastructure, dealing with the explosive growth in mobile data traffic.

A promising approach to deal with this phenomenon is to introduce caching at Base Stations (BSs). Content caching at BSs is indeed very beneficial for mobile operators for several reasons, the most important are: 1) it reduces the data traffic on the backhaul links, 2) it reduces the delay experienced by cellular network's users, and 3) it contributes in reducing the congestion during the peak hours.
Therefore, this issue has attracted the attention of the research and industrial communities.
A comprehensive survey on \textit{caching} in \textit{cellular networks}, and more specifically in 5G networks, along with its benefits and challenges is given in~\cite{wang2014cache}.

Furthermore, very recently, some ideas taken from network coding have been applied to caching in cellular networks~\cite{dimakis2010network,golrezaei2013femtocaching,altman2014distributed,avrachenkov2016optimization} and it was shown  that network coding-based caching policies further improve the performance obtained so far by classical caching schemes. The core idea of this technique is to use random linear network coding, where linear combination of contents (or chunks of files) are stored in the caches of BSs.

In this work we combine {\em linear content coding} techniques
and cellular network coverage models from {\em stochastic geometry},
to propose and evaluate some novel geographic caching policies that further improve the hit probability obtained in previous  works.
The idea is to store in the caches of base stations 
{\em linear combinations of contents} so as to increase the cache-hit probability by 
leveraging the probability of covering the request location by more
than one base station.
For a general content popularity distribution and 
a general distribution of the number of stations covering a typical
location in the network, we formulate a problem 
of the {\em optimal deterministic policy of content caching with network
coding}, maximizing the probability of serving the typical content
request from the caches of covering stations.
We find the solution to this problem using the {\em dynamic programming}
approach. We also consider two natural {\em greedy caching policies}.
Theoretical bounds can be given on the
sub-optimality of one of these greedy policies leveraging the 
classical theory of monotone sub-modular
set functions.

We evaluate numerically our policies considering two popular stochastic geometric 
coverage models: the Boolean one and the
Signal-to-Interference-and-Noise-Ratio (SINR) one,  and compare their
performance (the hit probability)  to those offered by the caching of
the most popular content in all base stations and an optimal random,
independent, caching strategy~from~\cite{blaszczyszyn2015optimal}, both considered  
as reference strategies in the literature.
Our numerical results show that the proposed policies employing
network coding are in general not worse than the two reference
policies and can further improve the total hit probability offered by
the independent caching policy in the moderately high coverage regime.


\subsection*{Related Work}
There is a considerable number of papers dealing with cellular
caching. In what follows we mention only the most relevant to our approach.

Bastug et al in~\cite{bacstug2015cache} provide some 
early stochastic geometry results on the user outage probability and average
delay experienced in cache-enabled  cellular networks, further developed in~\cite{bacstuug2016delay}.

The optimized independent caching policy, used here as a reference one, was proposed
in~\cite{blaszczyszyn2015optimal} and evaluated under the same 
Boolean and SINR coverage models.
Networks of wireless caches in the plane with geometric constraints
and linear network coding were first studied by Altman et al in~\cite{altman2014distributed}.

Avrachenkov et al in~\cite{avrachenkov2016optimization} considered an approach with linear network coding very similar to ours,
addressing the hit probability maximization as a  generalized unbounded
knapsack  problem. Having proved that the problem 
can be solved in general using dynamic programming, they mainly focus
on the explicit solution
for a special case when no coding is applied 
(each file is stored or not in a single chunk).
They also consider a relaxation of the problem in which
the average constraints on the storage space are imposed.
This latter model with one chunk happens to be equivalent to the one considered in~\cite{blaszczyszyn2015optimal}.
Our formulation of the hit maximization problem as a  monotone sub-modular
set function maximization, while sub-optimal 
with respect to the generalized knapsack problem
of~\cite{avrachenkov2016optimization},
cf Remark~\ref{second_remark}, has a two-fold advantage:
(1) the optimal (dynamic programming) solution has much simpler structure 
allowing one for efficient numerical evaluation of its performance in
a general case and (2) it admits natural approximations by greedy
policies. Both approaches prove to be competitive with respect to the
reference policies considered so far in the literature.

Further related works include~\cite{dai2012collaborative}, where
some auction-based collaborative caching mechanism for wireless streaming is studied.
In \cite{gharaibeh2016aprovably}, the authors studied the on-line
collaborative caching problem for a multicell-coordinated system from
the point of view of minimizing the total cost paid by the Content
Providers.
Similar scenarios are considered in~\cite{li2015delay,wang2015framework}.
Cooperative caching and cooperative redundancy elimination models were proposed in~\cite{wang2013intra} for an intra-Autonomous System.

The remaining part  of the paper is organized as follows.
Section~\ref{model_notations} introduces the caching and network coding as well as the user coverage models used in the paper.
Section~\ref{opt_models} describes the optimal and greedy caching policies that we propose to solve the hit probability maximization problem.
Evaluation of the proposed policies under the considered coverage models is done in Section~\ref{numerical_results}.
Finally, concluding
remarks and future research are discussed in Section~\ref{conc}.


\section{Caching with Network Coding Model}
\label{model_notations}

\subsection{Network Coverage Distribution}
We consider a general cellular-network germ-grain type~\cite[Section
  6.5]{chiu2013stochastic} coverage model generated by stationary marked point process $\Phi=\{(X_i,\cell_i)\}$,
where $X_i$ model the positions of Base Stations (BSs) and
$X_i\oplus\cell_i$ their cells (service zones)~\footnote{$\cell_i$ are assumed
to be random closed sets and $X_i\oplus\cell_i$ means the translation
of the set $\cell_i$ by $X_i$.}.
In fact, the only characteristics of this model, used in the remaining part of this paper, is the  {\em coverage (number) distribution}  defined as 
$$p_k:=\Pr \{\,\coverageN=k\,\},\qquad k=0,1,\ldots,$$
where $\coverageN:=\sum_{i}\ind(0\in X_i\oplus\cell_i)$ is the number of cells covering the 
typical location, assumed without loss of generality (due to stationarity of the model)
to be the origin~0 of the plane. 
We consider $\coverageD:=\{p_k:k=0,1,\ldots\}$ as a (given) probability distribution on $\coverageset:=\{0,1,\ldots\}$.
We shall use also the following notation for 
the tail-distribution function of the 
 coverage number 
$$\coverageTDF(k):=\Pr\{N\ge k\}=\sum_{n=k}^\infty p_n.$$

\subsubsection{Boolean Model}
\label{bl_cvg_model}
An important special case of the coverage model considered in the
literature and in this paper is  the {\em Boolean Model} (BM), where
$\Phi$ is a homogeneous Poisson point process and $\{\cell_i\}$ are independent, identically distributed closed sets of finite mean surface $\Ex[|\cell_i|]<\infty$.
In this case the coverage number $\coverageN$ has a Poisson distribution
with parameter $\lambda':=\lambda\Ex[|\cell_i|]$, where $\lambda>0$ is the intensity of  $\Phi$ (corresponding to the density of BSs); cf~\cite[Lemma~3.1]{FnT1}.
Consequently, for the BM we have 
\begin{equation}\label{e.BMcoverage}
p_k=e^{\lambda'}\frac{{\lambda'}^k}{k!},\qquad k=0,1,\ldots
\end{equation}

\subsubsection{SINR model}
\label{sinr_cvg_model}
In a  more adequate model for cellular network, called {\em Signal-to-Interference-plus-Noise Ratio (SINR) coverage model}~\footnote{Also  called  shot-noise germ-grain model in~\cite[Section~6.5.4]{chiu2013stochastic}.}, the coverage number distribution is given by a more complicated expression; cf~\cite{blaszczyszyn2015studying}. For this model we have
\begin{equation}\label{e.SINRcoverage}
p_k:=\sum_{n=k}^{\lceil 1/\tau\rceil}  
(-1)^{n-k}{n\choose  k} \mathcal{S}_n(\tau) \,,
\end{equation}
where
\begin{equation}\label{e.SINRcoverageNb}
\mathcal{S}_n(\tau) = \tau_n^{-2n/\beta}  \calI_{n,\beta}((W)a^{-\beta/2}) \calJ_{n,\beta}(\tau_n)
\end{equation}
%
represents the expected number of $n$-tuples of BSs the typical user
can select  among those which cover it with the  SINR greater than
$\tau$. The notation used in the above expressions is as follows:  $\tau_n:=\frac{\tau}{1-(n-1)\tau}$,
$\beta$ is the path-loss exponent\footnote{The path-loss function is
  $(Kr)^{-\beta}$, with constants $K>0$ and $\beta >2$, and $r$ is the
  distance between the BS and the user.}, $W$ is the external noise
power,  and the two special functions 
\begin{equation}\label{In}
\calI_{n,\beta}(x):=\frac{2^n
\int\limits_0^{\infty} u^{2n-1}e^{-u^2-u^\beta x\Gamma(1-2/\beta)^{-\beta/2}} du
}{\beta^{n-1}\Gamma(1-2/\beta)^n\Gamma(1+2/\beta)^n(n-1)!}
\end{equation}
with
$a=\frac{\lambda\pi
  \Ex[(PS)^{\frac{2}{\beta}}]}{K^{2}}$, where $\lambda$ is the density
of BSs, $S$ is the fading/shadowing random variable, $P$ is the
BS transmission power, $K$ is the path-loss constant, and 
\begin{align}\nonumber
\calJ_{n,\beta}(x):= \frac{(1+nx) }{n}  
\hspace{-1.5em}\int\limits_{[0,1]^{n-1}}&\hspace{-1em}  \frac{  \prod_{i=1}^{n-1}
  v_i^{i(2/\beta+1)-1}(1-v_i)^{2/\beta}  }   {  \prod_{i=1}^n
  (x+\eta_i)}\\[-2ex]
&\hspace{6em}dv_1\dots dv_{n-1}, \label{Jn}
\end{align}
with 
$\eta_1= v_1v_2\dots v_{n-1}$,
$\eta_2= (1-v_1)v_2\dots v_{n-1}$, 
$\eta_3= (1-v_2)v_3\dots v_{n-1}$, $\ldots$, 
$\eta_n= 1- v_{n-1}$.
Note that in contrast to the Boolean model, the coverage number distribution in the SINR model has a bounded support; $N$ is 
not larger than the constant~$\lceil 1/\tau\rceil$
depending on the required SINR threshold~$\tau$.

\subsection{Content Popularity Distribution}
We consider a {\em finite}~\footnote{Finiteness of the content set 
simplifies the model analysis.}
 set of contents, indexed by a subset of  integers
 $\contentset:=\{1,2,\ldots,J\}$, $J<\infty$. Popularity of these contents is modeled
by a probability distribution $\{a_j:j\in\contentset\}$, called {\em (content) popularity distribution}.
The value $a_j$ is interpreted as the probability that a typical user (located at the origin~0) requests the content $\content=j$ 
from the network; $a_j:=\Pr\{\content=j\}$, where $\content$ is the
index of the requested content. Without loss of generality we assume
that the content items are indexed according to the decreasing
popularity: $a_1\ge a_2\ge\ldots\ge a_J$.
In what follows, we always assume that the requested content $\content$ and the coverage number $\coverageN$ are independent random variables.

An important special case of the content popularity distribution,
having some empirical justifications, is the truncated Zipf distribution, with
\begin{equation}\label{e.Zipf}
a_j = A^{-1} j^{-\gamma}, \qquad j=1\,\ldots\,J,
\end{equation} 
where $\gamma$ is the (Zipf) exponent and $A = \sum_{j=1}^{J}
j^{-\gamma}$.

\subsection{Content Placement and Recovery Using Network Coding}
We assume that a cache memory consisting of  $L\geq1$ {\em blocks}  is  available at each BS.
The size of  each block corresponds to the size of exactly  one content item (all content items are assumed to have the same size).
In this paper we assume that {\em all BSs store exactly the same subset of contents}.
The spatial diversity (leveraging multiple coverage) is achieved 
by using some network coding techniques in the contents storage implementation, allowing one to store  linear combinations of {\em more content items than the number of memory blocks}.

More precisely, in each block $i=1,\ldots,L$ a linear combination of the content items from $\cache_i\subset\contentset$ of cardinality $|\cache_i|:=\#(\cache_i)\ge1$ is stored.
All base stations encode in  their  memory blocks $i=1,\ldots,L$ 
exactly the same subsets $\cache_i$ of content items 
using {\em mutually independent linear combinations} of the contents.
Motivated by this, we assume that a user (say located at the origin)
requesting some  content item $j\in\contentset$ can effectively
recover it from the caches of the BSs covering it when $j$ is encoded
in some  block $\cache_i$ of contents of cardinality $|\cache_i|$ not
greater than the coverage number $N$, i.e., when
\begin{equation}\label{e.hit}
\min\{|\cache_i|: j\in\cache_i,i=1,\ldots,L\}\le N\,.
\end{equation}
When the condition~\eqref{e.hit} is satisfied, we say the requested content item $j$ {\em is hit} in the network.

Denoting by $\delta_j$, $j\in\contentset$, the indicator of the event~\eqref{e.hit} we can write the {\em hit probability}
of the content item $\content$ randomly selected according to the popularity distribution as a function of the 
choice of the subsets $\{\cache_i\}_{i=1}^L$ of contents encoded in
the memory blocks of BSs  as 
\begin{align}
\hitP=\hitP(\{\cache_i\})&:=\Pr\{\hit_I=1\}\nonumber\\
&=\sum_{j=1}^\infty a_j\coverageTDF\Bigl(\min\{|\C_i|: j\in\C_i\}\Bigr)\,,\label{e.hit-proba-1}
\end{align}
with $\min\{\emptyset\}=\infty$.


\section{Hit Probability Optimization Problem}
\label{opt_models}
By a {\em content caching policy} we mean in what follows a configuration of
$L$ sets of contents  $\{\C_i\}_{i=1}^L$ to be encoded in $L$ memory blocks of all BSs.
Our main goal in this section consists in finding a 
caching policy maximizing the hit probability, that is solving 
\begin{equation}\label{e.Problem1}
\max_{\C_1,\ldots,\C_L\subset\contentset}  \hitP(\{\C_i\})\,.
\end{equation}

We shall also present a few reasonable sub-optimal content caching policies.
Let us first remark the following
relations to some previously considered caching policies.
\begin{remark}
\label{first_remark}
Caching the $L$ {\em most popular contents} corresponds to  
taking $\C_j=\{j\}$, $j=1,\ldots,L$. This policy  is obviously the
  optimal one in the case of the {\em 1-coverage} regime ($p_k=0$
  for $k\ge 2$.).  
{\em Independent caching} proposed
  in~\cite{blaszczyszyn2015optimal} leverages multiple coverage to
  increase the hit probability, without using network coding.
In contrast to the policies considered
  in this paper, it is a {\em randomized policy} providing
  all BSs with some probability distribution on the set of content items
 (in fact the sequence of caching
probabilities for all contents) and letting BSs independently
sample the composition of their cached contents from this
distribution. This distribution  is calculated (as in the current
paper) in function of the content
popularities and BS coverage probabilities so as to maximize
the average (cache) hit probability. Note this  policy maximizes  the
hit probability but in a different class of policies and hence, in
general, one cannot easily compare it to policies considered in this paper.
\end{remark}
\begin{remark}
\label{second_remark}
Our current optimization problem is a restriction of the knapsack problem
stated  in~\cite{avrachenkov2016optimization}. The restriction comes from the fact that our policy $\{\C_i\}$ assumes coding content items separately for each memory block
$i=1,\ldots,L$ and using for all contents $j\in\C_i$ present in a given block~$i$ the same number of equations $n_i=M/|\C_i|$, where $M$ is the number of chunks of each item considered in~\cite{avrachenkov2016optimization}.
While this latter assumption is not restrictive ($n_i\not=M/n$ for some $n$ can
be shown sub-optimal), coding separately different memory blocks 
is indeed sub-optimal. This assumption however has an important
consequence: it transforms the generalized knapsack problem to a 
simpler sub-modular set function maximization problem, as will be shown in
what follows. 
\end{remark}

\subsection{Properties of the Optimal Caching Policies}
In what follows we present some properties satisfied by any policy maximizing~\eqref{e.Problem1}.

\begin{lemma}\label{l.properties}
There exists a policy $\{\C_i\}$ maximizing~\eqref{e.Problem1}
having the following properties:
$\bigcup_{i=1}^L \C_i=\{1,\ldots,j_{\max}\}$, 
$|\C_1|\le\ldots\le|\C_L|$,
and all elements of
$\C_i$ precede those of $\C_{i+1}$; i.e., 
$\C_i=\{|\C_1|\!+\!\ldots\!+\!|\C_{i-1}|\!+\!1,\!\ldots,\!|\C_1|\!+\!\ldots\!+\!|\C_{i-1}|\!+\!|\C_{i}|\}$.
\end{lemma}

\begin{IEEEproof}
Let $\{\C_i\}$ maximizing~\eqref{e.Problem1}.
Suppose an item $x\in\bigcup_{i=1}^L\C_i$ is present in more than one set $\C_i$. 
 Keeping $x$ only in one set of smallest cardinality 
does not decrease the hit probability. Assume hence that $\{\C_i\}$ are
pairwise disjoint and 
suppose there exist  $x\in \dot{\bigcup}_{i=1}^{L} \C_i$ and 
$y\not\in \dot{\bigcup}_{i=1}^{L} \C_i$, such that $y<x$. 
Replacing $x$ by $y$ 
does not decrease  the hit probability.
Consider hence the case $\dot\bigcup_{i=1}^L \C_i=\{1,\ldots,j_{\max}\}$.
Assume now that $\{\C_i\}$ is indexed in the increasing order of cardinalities.
Suppose that $x \in\C_i$, $y \in\C_j$ with $|\C_i|\le|\C_j|$ and $x>y$.
Then, swapping $x$ and $y$ does not decrease the hit probability and
we can construct a new partition $\{\C_i\}$ of $\{1,\ldots,j_{\max}\}$
in which  all elements of
$\C_i$ precede those of $\C_{i+1}$.
\end{IEEEproof}

Note that a policy $\{\C_i\}$ satisfying the conditions of
Lemma~\ref{l.properties} has pairwise disjoint sets $\C_i$. It is easy to see that this simplifies the
expression~\eqref{e.hit-proba-1} of the hit probability to the following one. 
\begin{lemma}\label{l.hit-proba-2}
For pairwise disjoint $\C_i$, $i=1,\ldots,L$. 
\begin{equation}\label{e.hit-proba-2}
\hitP(\{\C_i\})=
\sum_{k=1}^{L} \popularity(\C_k)\coverageTDF(|\C_k|)\,.
\end{equation}
\end{lemma}

In view of~Lemma~\ref{l.properties} we can simplify the problem~\eqref{e.Problem1} restricting ourselves to the policies of the form $\C_i=[m_1\!+\!\ldots\!+\!m_{i-1}\!+\!1,m_1\!+\!\ldots\!+\!m_{i}]$, where
$[k,l]:=\{k,k+1,\ldots,l\}$ for integers $k,l$. By Lemma~\ref{l.properties} and Lemma~\ref{l.hit-proba-2}:
\begin{proposition}\label{p.the-problem-2}
We have
\begin{align}\label{e.the-problem-2}
&\max_{\C_1,\ldots,\C_L\subset\contentset}  \hitP(\{\C_i\})\\
&\!\!\!=\hspace{-2ex}
\max_{1\le m_1\le\ldots\le m_L}\!
\sum_{k=1}^{L}\!\!\popularity([m_1\!+\!\ldots\!+\!m_{k-1}\!+\!1,m_1\!+\!\ldots\!+\!m_k])\coverageTDF(m_k)\,.
\nonumber
\end{align}
\end{proposition}

\subsection{Dynamic Programming Solution of the Optimal Caching Problem}
\label{ss.DP}
The idea  consists in finding first the optimal cardinality  $m_L$ of the last block $\C_L$
in function of the assumed (unknown beforehand) total number $n$ of contents cached in previous blocks.
Then proceed recursively with cardinalities $m_l$ of blocks $\C_l$, $2\le l \le L-1$,  maximizing them in function of the assumed total number $n$ of contents  cached in $\C_1,\ldots,\C_{l-1}$ while taking into account already calculated contribution of the blocks $l+1,\ldots,L$. This leads to the following Dynamic Program (DP):

We express the optimal number of contents cached in the $L$-th
block as a function of the total number $n\in\contentset$ of contents cached in previous blocks
and the corresponding hit probability as:
\begin{align*}
m_L(n)&:= \argmax_{x}\popularity([n+1,n+x])
\coverageTDF(x),\\ 
\hitP(L,n)&:= \max\limits_{x}
\popularity([n+1,n+x])
\coverageTDF(x).
\end{align*}
By the induction, for a block~$l$, with $2 \leq l \leq L-1$, we
define:
\begin{align*}
&m_{l}(n):= \argmax_x \Bigl(\!\popularity([n+1,n+x])\coverageTDF(x)+\hitP(l+1,n+x)\!\Bigr),\\
&\hitP(l,n):= \max_x \Bigl(\!\popularity([n+1,n+x])+ \hitP(l+1,n+x)\!\Bigr).
\end{align*}
Finally, for the first block  we consider
only  $n=0$ since in this case we should start with the first content item
\begin{align*}
m_{1}&:= \argmax_x \Bigl(\popularity([1,x])\coverageTDF(x)+\hitP(2,x)\Bigr),\\
\hitP(1)&:= \max_x \Bigl(\popularity([1,x])+ \hitP(2,x) \Bigr).
\end{align*}
The above DP approach leads to the following solution of our optimal caching problem~\eqref{e.the-problem-2}.
\begin{proposition}
The maximal hit
probability in~\eqref{e.the-problem-2} is equal to $\hitP(1)$  and it
is achieved on $(m^*_1,\ldots,m^*_L)$ defined as 
\begin{equation}\label{e.m^*}
\begin{split}
m^*_1&:=m_1\\
m^*_2&:=m_{2}(m^*_1)\\
&\ldots\\
m^*_{L-1}&:=m_{L-1}(m^*_{1}+m^*_{2}+\ldots+m^*_{L-2})\\
m^*_{L}&:=m_{L}(m^*_{1}+m^*_{2}+\ldots+m^*_{L-1})\,.
\end{split}
\end{equation}
\end{proposition}

\subsection{Greedy Sub-Optimal Caching Policies}
The DP approach to the hit probability maximization~\eqref{e.Problem1} completely solves the problem but 
presents a considerable numerical complexity. 
Greedy algorithms are supposed to propose simpler policies, reasonably approaching the maximal hit probability~\eqref{e.Problem1}.
Depending on whether we apply the greedy approach to the class of general policies $\{\C_i\}$, using the expression~\eqref{e.hit-proba-1} as the definition of the hit probability, or we restrict ourselves to the class of structured policies $\C_i=[m_1+\ldots+m_{i-1}+1,m_1+\ldots+m_i]$, using the expression~\eqref{e.hit-proba-2} for this probability, we obtain two, in general different, greedy  policies, both in general suboptimal. The former one, using general $\{\C_i\}$,
has interesting theoretical bounds regarding its sub-optimality, but still represents considerable numerical complexity.
The latter one, assuming the structured policies, is numerically much simpler, but we do not have any theoretical guarantees regarding its performance. Numerical evidences suggest its utility.
Finally, note that, even if the optimal policy is known to have the structured form, a greedy algorithm operating in this class is in general worse than the greedy algorithm operating in the set of all non-structured policies.

\subsubsection{Greedy Caching Policy with General Blocks}
\label{sss.GGB}
Let us choose the first set $\C^g_1$ of items as a subset maximizing the one-block hit probability $\hitP(\{\C\})$.
Since $a_i$ are  decreasing $\C^g_1$ has the form $\C^g_1=[1,\ldots,m^g_1]$ for some $m^g_1\ge1$.
Thus
\begin{equation}\label{e.one-block-hit-proba}
\C^g_1\in\argmax_{\C\subset\contentset} \popularity(\C)\coverageTDF(|\C|)=
\argmax_{m_1\ge1}\popularity([1,m_1])\coverageTDF(m_1).
\end{equation}
Then recursively, let us choose sets $\C^g_l$, $2\le l\le L$ 
maximizing the increment of the hit probability they offer,
{\em without assuming mutual disjointness of the sets}
\begin{align}\nonumber
&\C^g_l\in \argmax_{\C\subset\contentset} \hitP(\{\C^g_1,\ldots,\C^g_{l-1},\C\})-\hitP(\{\C^g_1,\ldots,\C^g_{l-1}\})\\
&=\argmax_{\C\subset\contentset}\!\! 
\sum_{j\in \C}^\infty \!\!a_j \Bigl(\coverageTDF(|\C|)-
                            \coverageTDF\bigl(\min\{|\C^g_i|:
                            j\in\C^g_i,i\le l-1\}\bigr)\Bigr)^+\,,
\label{e.l-block-hit-proba-increase-general}
\end{align}
where $(x)^+=\max(x,0)$.

The following result not only gives a lower bound on the hit probability achieved by $\{C^g_i\}$
but also allows one to mitigate the decrease  of this  probability by increasing the number of memory blocks.  
\begin{proposition}\label{p.greedy-general}
Let $\{\C^g_i\}_{i=1}^{K}$ be  contents sets selected by the  greedy caching policy~\eqref{e.one-block-hit-proba}, \eqref{e.l-block-hit-proba-increase-general} applied for  $K\ge L$ memory blocks
and $\{C^*_i\}_{i=1}^L$ an optimal solution of the problem~\eqref{e.Problem1} for $L$ memory blocks.
Then
\begin{equation}\label{e.greedy-general}
\hitP(\{C^g_i\}_{i=1}^K)\ge (1-e^{-L/K})\hitP(\{C^*_i\}_{i=1}^L)\,.
\end{equation}
\end{proposition}
\begin{IEEEproof}
Consider $\hitP(\cdot)$ defined by~\eqref{e.hit-proba-1}
as a set function on the space of finite subsets $\{\C_1,\ldots,\C_i\}$ of finite subsets  $\C_i\subset\contentset$, $i\ge1$.
Clearly $\hitP(\cdot)$ is non-negative, increasing 
$$ \hitP(\{C_1,\ldots,\C_i\})\le\hitP(\{C_1,\ldots,\C_i,\C_{i+1},\ldots,\C_{i+k}\})$$
and sub-modular
\begin{align*}
&\hitP(\{C,C_1,\ldots,\C_i\})-\hitP(\{C_1,\ldots,\C_i\})\\
\ge&
\hitP(\{C,C_1,\ldots,\C_i,\C_{i+1},\ldots,\C_{i+k}\})-\\
& \hitP(\{C_1,\ldots,\C_i,\C_{i+1},\ldots,\C_{i+k}\})\,.
\end{align*}

Indeed, for this latter property observe that the right hand side of~\eqref{e.l-block-hit-proba-increase-general} is decreasing with respect to~$l$.
The result follows thus by the classical result~\cite{Nemhauser1978} for sub-modular set functions.
\end{IEEEproof}

\subsubsection{Greedy Caching Policy with Disjoint Blocks}
\label{sss.GDB}
The first set $\C^{gd}_1:=\C^g_1=[1,m^g_1]$ is chosen by this policy as for the previous greedy policy~\eqref{e.one-block-hit-proba}.
Then recursively, let us choose sets $\C^{gd}_l=[m^g_1+m^g_2+\ldots+m^g_{l-1}+1,m^g_1+m^g_2+\ldots,m^g_{l}] $, $2\le l\le L$ 
maximizing the increment of the hit probability they offer,
{\em assuming mutual disjointness of the sets}, thus using expression~\eqref{e.hit-proba-2} for this probability
\begin{align}\nonumber
&\C^{gd}_l\in \argmax_{\C\subset\contentset} \hitP(\{\C^g_1,\ldots,\C^g_{l-1},\C\})-\hitP(\{\C^g_1,\ldots,\C^g_{l-1}\})\\
&=\hspace{-1em}\argmax_{m_l\ge m_1+\ldots+m_{l-1}}\hspace{-1em}\popularity([m_1\!+\!\ldots\!+\!m_{l-1}+1,m_1\!+\!\ldots\!+\!m_l])\coverageTDF(m_l).
\label{e.l-block-hit-proba-increase-disjoint}
\end{align}


\section{Numerical Results}
\label{numerical_results}

In this section, we evaluate the performance of the proposed caching policies described previously considering the BM and SINR coverage models.
In particular we  study the dependence of our policies on the
mean number of stations $\Ex[N]$ covering a given location
and the Zipf exponent~$\gamma$ of the content popularity distribution.
We first describe the numerical setup, including adopted coverage models (Sec.~\ref{subsection_numerical_setup}), and then we analyze and discuss the numerical results (Sec.~\ref{subsection_performance_evaluation}).

\subsection{Numerical Setup}
\label{subsection_numerical_setup}

Let us first give more details about the coverage models used in our results.

\subsubsection{Boolean Model}
We assume that the interference is small compared to noise (\emph{noise-limited case}) and hence we use the Boolean model to calculate the probability of user coverage by $k$~BSs ($p_k$ in~\eqref{e.BMcoverage}).
The signal-to-noise ratio can be expressed as $\frac{P (K r)^{-\beta}}{W}$, where $P$ is the BS transmit power, $K$ is the path-loss constant, $r$ is the distance between the BS and the user, $\beta$ is the path-loss exponent and $W$ is the noise power.
Let $B=K(\frac{P}{W})^{1/\beta}$.
Hence, we have $\Ex[|\cell_i|] =  \pi \tau^{-2/\beta} B^{-2}$ and
$\lambda' = \lambda \pi \tau^{-2/\beta} B^{-2}$ (see
Section~\ref{bl_cvg_model}). 
Note, the mean coverage number is equal to $\Ex[N] =\pi \lambda \tau^{-2/\beta} B^{-2}$. 


\subsubsection{Signal-to-Interference Ratio (SIR) Model}
For general shadowing conditions, the coverage probability $p_k$ is calculated in~\eqref{e.SINRcoverage}, Section~\ref{sinr_cvg_model}.
We use a package developed in Matlab, available at~\cite{codeMatlabkeeler2014}, to compute the numerical values of the probabilities $p_k$ for this model.
The mean coverage is equal to $\Ex[N] = \mathcal{S}_1(\tau)$.

To evaluate the effectiveness of the proposed content caching policies, we conduct calculations using Matlab.
Default values of key numerical parameters are as follows:
density of BSs $\lambda=1$, number of cache blocks $L=5$, total number of
content items $J=40$, Zipf exponent $\gamma\in\{0.9,0.56\}$, path-loss exponent $\beta=3$ and constant $K=1$, noise power $W=0$ in SIR
model and $P/W=1$ in BM.
%
\begin{figure}[t!]
	\centering

	\subfloat[$L=5$, $\lambda=1$, $\gamma=0.9$, Boolean model]
	{
		\includegraphics[width=0.5\textwidth]{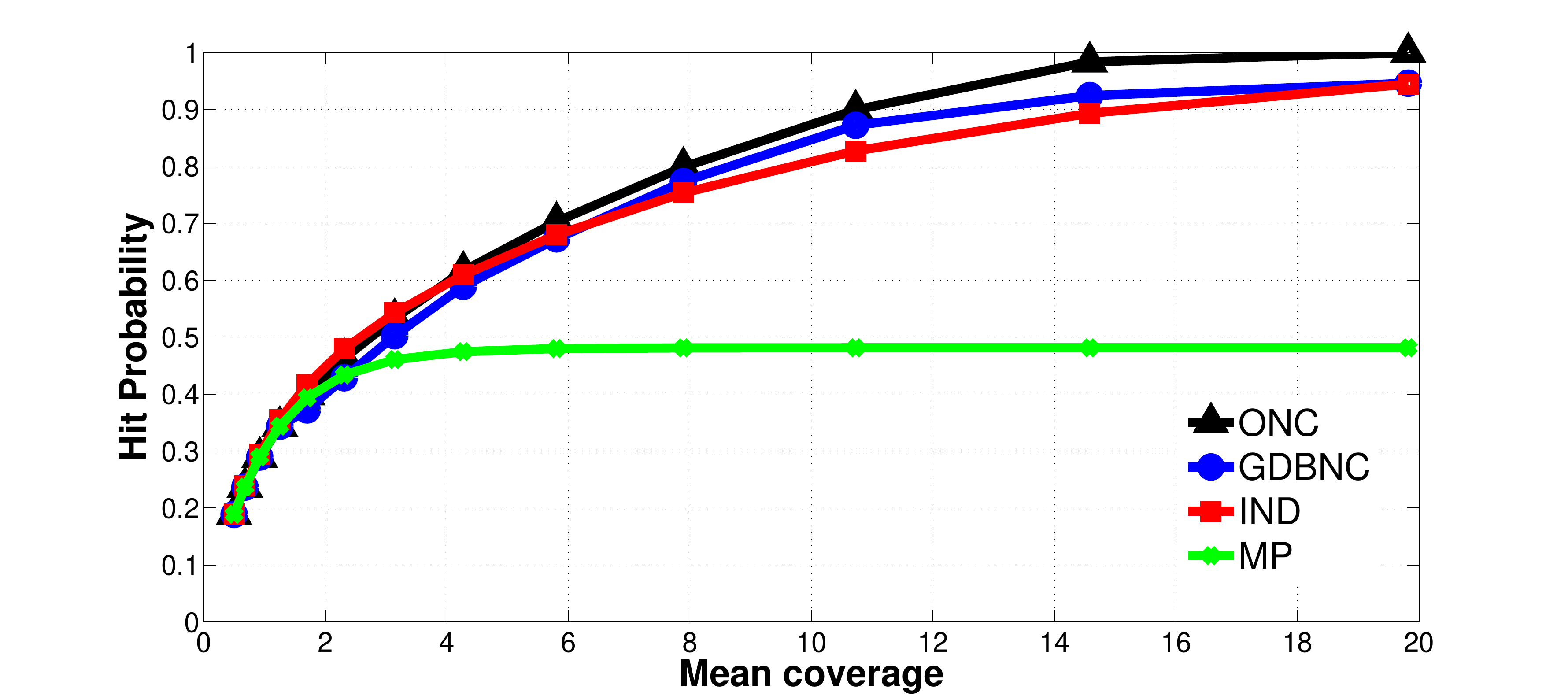}
		\label{bool_hit_prob_L5_lambda1gamma09_meanCov}
	} \\ 
 	\subfloat[$L=5$, $\lambda=1$, $\gamma=0.56$, Boolean model]
	{
		\includegraphics[width=0.5\textwidth]{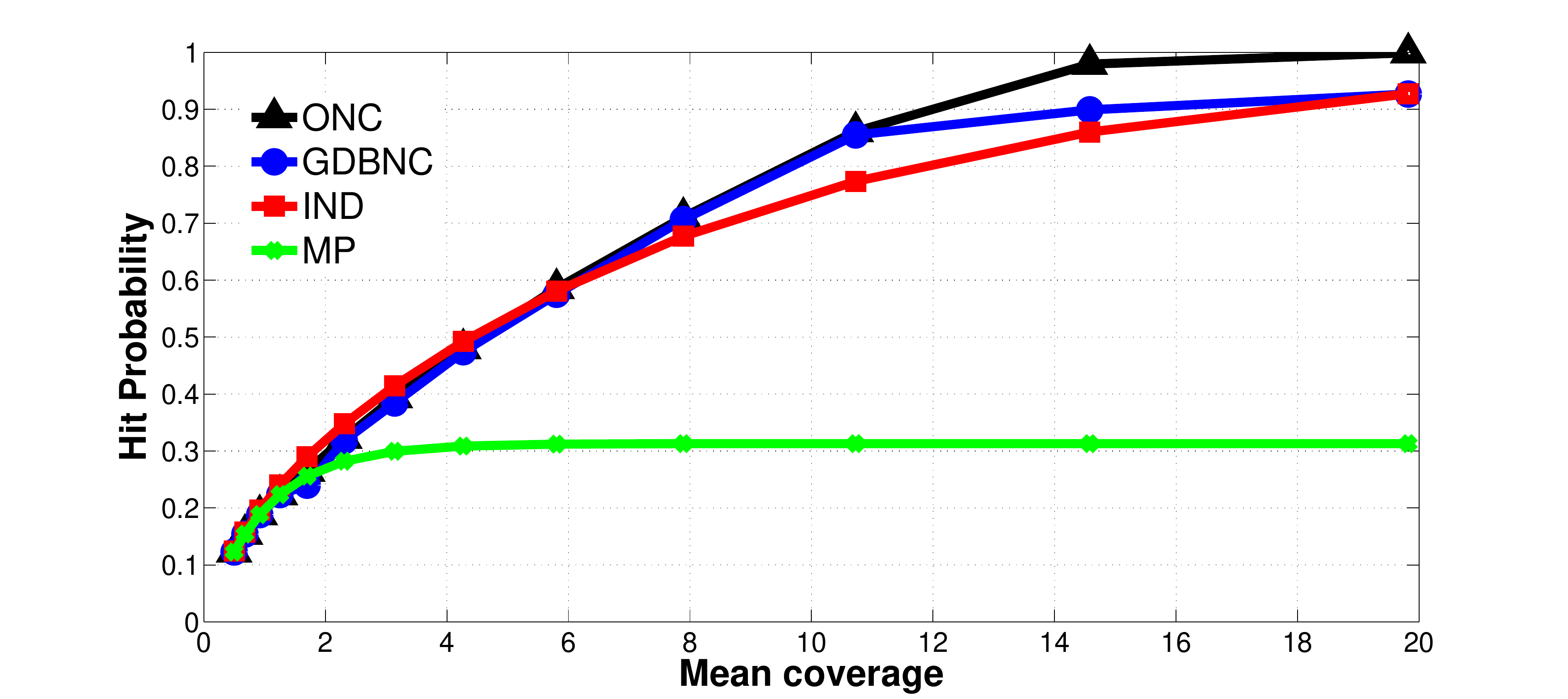}
		\label{bool_hit_prob_L5_lambda1gamma056_meanCov}
	}\\
	 \subfloat[$L=5$, $\lambda=1$, $\gamma=0.9$, SINR model]
	{
		\includegraphics[width=0.5\textwidth]{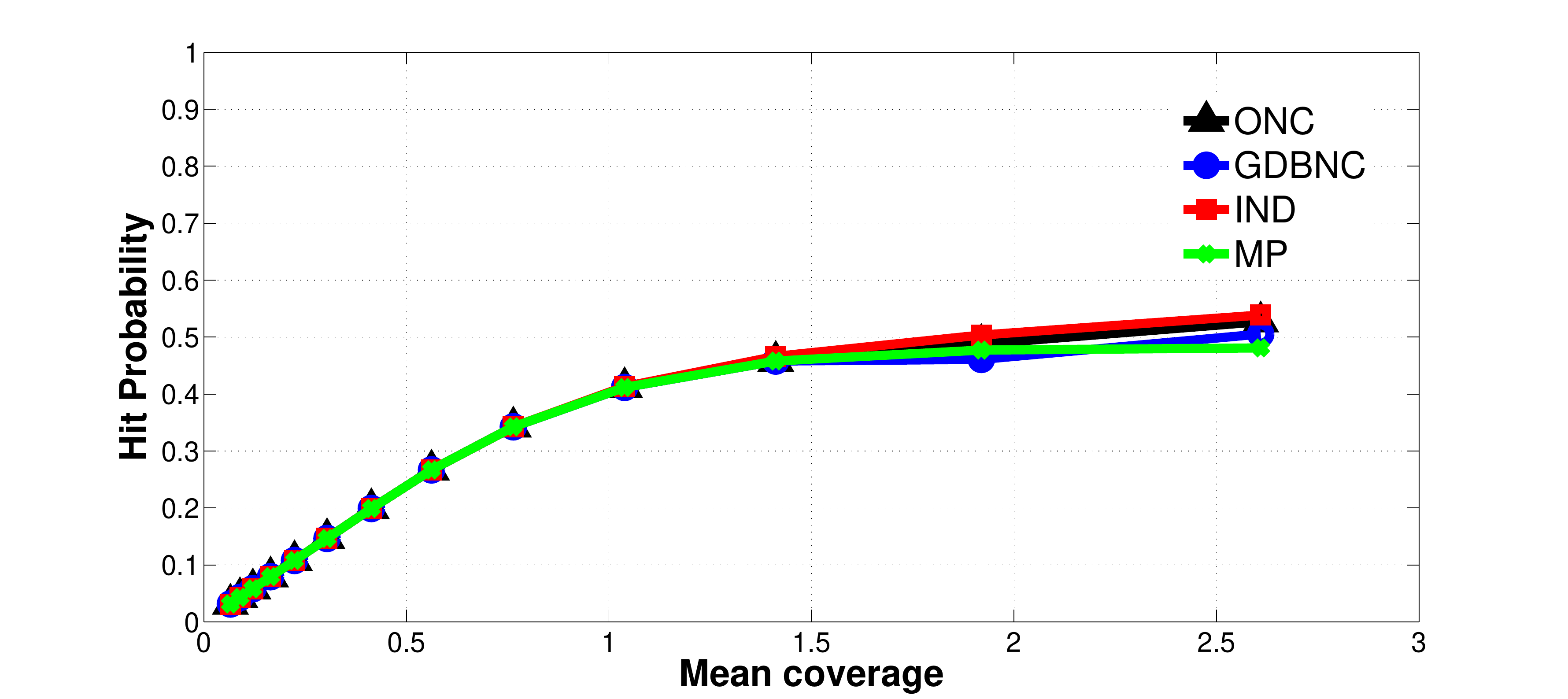}
		\label{sinr_hit_prob_L5_lambda1gamma09_meanCov}
	}
	\vspace{-1ex}
	\caption{\small{Hit probability $\hitP$ versus the mean coverage for $\tau \in$~[-12~dB, 12~dB], cache size $L=5$, $\lambda=1$ and $\gamma \in \{0.9,0.56\}$.}}
	\label{hit_prob_L5_lambda1gamma09056}
	\vspace{-1em}
\end{figure}
%
\subsection{Performance Evaluation}
\label{subsection_performance_evaluation}

We evaluate the optimal policy considered in Section~\ref{ss.DP},
called in what follows the  Optimal Network Coding (ONC) policy, 
and the greedy policy proposed in Section~\ref{sss.GDB} called
the Greedy Disjoint-Blocks Network Coding (GDBNC) policy.
We plot  the hit probability $\hitP$ versus  the \emph{mean coverage} $\Ex[N]$ for the BM and the SINR models.

We further compare our policies to the Most Popular (MP) and
Independent (IND) caching policy
discussed in  Remark~\ref{first_remark}. 

The results are discussed in the next subsections.

\subsubsection{Hit probability under the Boolean model}~\\

Figure\ref{hit_prob_L5_lambda1gamma09056}
\subref{bool_hit_prob_L5_lambda1gamma09_meanCov} and Figure\ref{hit_prob_L5_lambda1gamma09056}
\subref{bool_hit_prob_L5_lambda1gamma056_meanCov}
show the total hit probability versus the mean coverage, when $\tau$ varies in the range~[-12~dB, 12~dB], $L=5$, $\lambda=1$, and for $\gamma=0.9$ and $0.56$, respectively.
It can be observed that both the optimal and the greedy policies give
us a very good performance compared to MP. This is especially true for
a mean coverage value higher than 2.
For small mean coverage values (or equivalently for high $\tau$
values), all considered policies perform similarly to MP.

Furthermore, ONC performs better than IND for larger mean
coverage values. The gain is more important when the content
popularity distribution is more flat (smaller $\gamma$). 
The greedy policy GDBNC is close to the optimal ONC in some
intermediate coverage regime and joins IND when the coverage
ultimately increases.
It is an open question whether the same holds true for the greedy
policy with general bocks considered in  Section~\ref{sss.GGB}, which
is in between  GDBNC and ONC.

\subsubsection{Hit probability under the SINR  model}~\\
We now evaluate our policies under the SINR model.
In this case, the number of BSs covering the user is very much limited (i.e., $\Ex[N] < 3$).
As before, we plot in
Figure\ref{hit_prob_L5_lambda1gamma09056}\subref{sinr_hit_prob_L5_lambda1gamma09_meanCov}
$\hitP$ versus the \emph{mean coverage}. So, it is not surprising to
obtain the same behavior under all policies, except for mean coverage
bigger than $1.5$.
In fact, in this latter case, some performance gains could be obtained
by our policies with respect to the MP policy. However, for  the considered
parameters of the SINR model, there is no much enough room to improve 
$\hitP$ with respect to MP.

\section{Conclusion}
\label{conc}

In this paper we show how network coding ideas can be used to improve 
the performance of caching in cellular networks.
Specifically, we study a geographic caching problem in cellular
networks allowing for linear content coding at base stations.
Three policies are proposed: an optimal for our model and two natural
greedy ones. We show that all considered policies are equivalent to
caching in every base station the most popular content (without any coding) when there is no
coverage diversity.
However, when the number of stations covering a typical user increases, 
our policies perform much better than this trivial policy and
sometimes even better, in terms of the hit rate,  than the randomized caching
policy~\cite{blaszczyszyn2015optimal} previously proposed for
such regimes.

\end{document}